\documentclass[twocolumn]{aastex63}
\usepackage{natbib}
\usepackage{threeparttable}

\usepackage{makecell}
\usepackage{amsmath} % or simply amstext

\usepackage{float}
\usepackage{textcomp}

\usepackage{lineno}
\accepted{\today}
\submitjournal{ApJ Letters}
\shorttitle{LMC Bow Shock}

\shortauthors{Setton et al.}

\begin{document}

\title{The Large Magellanic Cloud's $\sim30$ Kiloparsec Bow Shock and its Impact on the Circumgalactic Medium}

\author[0000-0003-4075-7393]{David J. Setton }\thanks{E-mail: davidsetton@princeton.edu}
\thanks{Brinson Prize Fellow}\affiliation{Department of Astrophysical Sciences, Princeton University, 4 Ivy Lane, Princeton, NJ 08544, USA}
\affiliation{Department of Physics and Astronomy and PITT PACC, University of Pittsburgh, Pittsburgh, PA 15260, USA}

\author[0000-0003-0715-2173]{Gurtina Besla}
\affiliation{Steward Observatory, University of Arizona, 
Tucson, AZ 85705}

\author[0000-0002-9820-1219]{Ekta~Patel}\thanks{Hubble Fellow}
\affiliation{Department of Astronomy, University of California, Berkeley, 501 Campbell Hall, Berkeley, CA, 94720, USA}
\affiliation{Department of Physics and Astronomy, University of Utah, 115 South 1400 East, Salt Lake City, Utah 84112, USA}

\author[0000-0002-3817-8133]{Cameron Hummels}
\affiliation{TAPIR, California Institute of Technology, Pasadena, CA 91125, USA}

\author[0000-0003-4158-5116]{Yong Zheng}
\affiliation{Department of Physics, Applied Physics and Astronomy, Rensselaer Polytechnic Institute, Troy, NY 12180}

\author[0000-0001-9735-7484]{Evan Schneider}
\affiliation{Department of Physics and Astronomy and PITT PACC, University of Pittsburgh, Pittsburgh, PA 15260, USA}

\author[0000-0002-0197-526X]{Munier Salem}
\affiliation{Stripe, 354 Oyster Point Blvd South San Francisco, 94080}

\begin{abstract}

The interaction between the supersonic motion of the Large Magellanic Cloud (LMC) and the Circumgalactic Medium (CGM) is expected to result in a bow shock that leads the LMC's gaseous disk. In this letter, we use hydrodynamic simulations of the LMC's recent infall to predict the extent of this shock and its effect on the Milky Way's (MW) CGM. The simulations clearly predict the existence of an asymmetric shock with a present day stand-off radius of $\sim6.7$ kpc and a transverse diameter of $\sim30$ kpc. Over the past 500 Myr, $\sim8\%$ of the MW's CGM in the southern hemisphere should have interacted with the shock front. This interaction may have had the effect of smoothing over inhomogeneities and increasing mixing in the MW CGM. We find observational evidence of the existence of the bow shock in recent $H\alpha$ maps of the LMC, providing a potential explanation for the envelope of ionized gas surrounding the LMC. Furthermore, the interaction of the bow shock with the MW CGM may also explain observations of ionized gas surrounding the Magellanic Stream.  Using recent orbital histories of MW satellites, we find that many satellites have likely interacted with the LMC shock. Additionally, the dwarf galaxy Ret2 is currently sitting inside the shock, which may impact the interpretation of reported gamma ray excess in Ret2. This work highlights how bow shocks associated with infalling satellites are an under-explored, yet potentially very important dynamical mixing process in the circumgalactic and intracluster media.
\end{abstract}


\section{Introduction} \label{sec:intro}

The Circumgalactic Medium (CGM) plays a crucial role in the evolution of galaxies \cite[e.g.,][]{Tumlinson2011, Putman2012, Tumlinson2017, Peroux2020}. An understanding of this multi-phase medium is essential to modeling the mechanisms that replenish and deplete the gas reservoir in the star forming interstellar medium over a galaxy's lifetime. The CGM is a massive reservoir containing enriched material produced in star formation and subsequent supernovae explosion \cite[e.g.,][]{Peeples2014}. However, there is still considerable uncertainty in our understanding of the baryon cycle in galaxies, especially because there is strong evidence that the CGM at $z=0$ is still out of hydrostatic equilibrium \cite[e.g.,][]{Werk2014, Conroy2021,Lochhaas2023}. As such, understanding the multi-phase structure of the CGM and the sources of mixing is necessary to understand the flow of gas in and out of the CGM. 

Much of the focus on the state of the CGM has been on inflows and outflows as the predominant mechanisms that influence its multi-phase structure. Simulations predict that at z=0 the CGM of galaxies is still being supplied with low-metallicity gas from the intergalactic medium \citep[e.g.,][]{Stern2020}. Outflows of the enriched galactic interstellar medium have been observed in galaxies, driven by star formation \citep[e.g.,][]{Tremonti2007, Rubin2014, Diamond-Stanic2021} and active galactic nuclei, which may also serve to mechanically heat the existing CGM \citep[see references in][]{Alexander2012, Kormendy2013}. Additionally, the accretion of cool CGM gas enriches a galaxy's interstellar medium, as has been observed in star forming galaxies \citep[e.g.,][]{Rubin2012}.

While much of the observational work on the CGM has focused on gas entering via either outflows from the host galaxy or accretion from the IGM, simulations have shown that in MW mass halos as much as $\sim20\%$ of the gas mass of the CGM is expected to have originated in satellite galaxies \citep{Hafen2019}. In this study, we examine the impact of satellite galaxies on the structure and dynamics of the host galaxy's CGM, rather than its contribution to the mass budget. 

Cosmological simulations show that the dominant building block of MW mass halos ($\sim 10^{12}$ M$_\odot$) are subhalos that harbor $\sim$10\% of the host mass \citep{Stewart2008}. Indeed, the MW's most massive satellite galaxy, the Large Magellanic Cloud (LMC), is believed to have had a halo mass of $M_\mathrm{halo} \sim 10^{11}$ M$_\odot$ at infall \citep{Besla2012}. Thus, we consider the specific case of the LMC, which is just past its pericentric approach to the MW \citep{Besla2010, Kallivayalil2013}. Consequently, the LMC is moving very quickly, at 320 km/s relative to the MW \citep{Kallivayalil2013}. This speed implies that the LMC is moving supersonically through the MW's CGM and should generate a bow shock. It has been previously suggested that the LMC hosts a bow shock that may impact its star formation history \citep{deBoer1998}. Here, we use detailed hydrodynamical simulations from \cite{Salem2015} to predict the existence, morphology and kinematics of the LMC's bow shock using the known 3D velocity vector of the LMC. 

There is clear evidence for interaction between the LMC and the CGM of the MW in the form of the Magellanic Stream, a gas structure that trails behind the Magellanic Clouds \citep{Mathewson1974}. The origin of the Stream has been attributed to a combination of tidal forces and ram pressure stripping of the LMC's interstellar medium \citep{donghia2016}. Ram pressure stripping will result in the truncation of the LMC's gas disk. Indeed its gas disk is observed to be significantly smaller than its stellar disk (radius of 6 kpc vs. $\sim$ 18 kpc, see \citealt{Mackey2018}). Hydrodynamic simulations of the impact of ram pressure on the LMC's gaseous disk have reproduced the observed truncation of the LMC's gas disk \citep{Salem2015}, proving that the LMC is hydrodynamically interacting with the MW's CGM. 

In this letter, we further examine the \cite{Salem2015} simulations to study the impact of the supersonic motion of the LMC on the MW CGM itself. We posit that the resulting bow shock plays a key role in mixing the CGM as the $\sim30$-kpc shock swept through the southern hemisphere during the LMC's first infall. We predict the shape, jump conditions, and physical extent of this shock and discuss the consequences of this structure to our understanding of the CGM and relationship with dwarf satellites of the Milky Way. 

This letter is laid out as follows. In Section \ref{sec:sim}, we describe the updated run of the fiducial simulation from \cite{Salem2015} that we use in this work. In Section \ref{sec:shape}, we present our predictions for the physical properties of the LMC bow shock. In Section \ref{sec:observability}, we estimate the observability of this shock. Finally, in Section \ref{sec:discussion}, we discuss the influence that the LMC's bow shock may have had on: 1) mixing of the MW's CGM; 2) the ISM of several of Milky Way's other dwarf satellite galaxies; and 3) the interpretation of observed ionized gas associated with the Magellanic System. Unless otherwise specified, throughout this work ``CGM" refers to the circumgalactic medium of the Milky Way.

\begin{figure*}
\centering
\includegraphics[width=\textwidth]{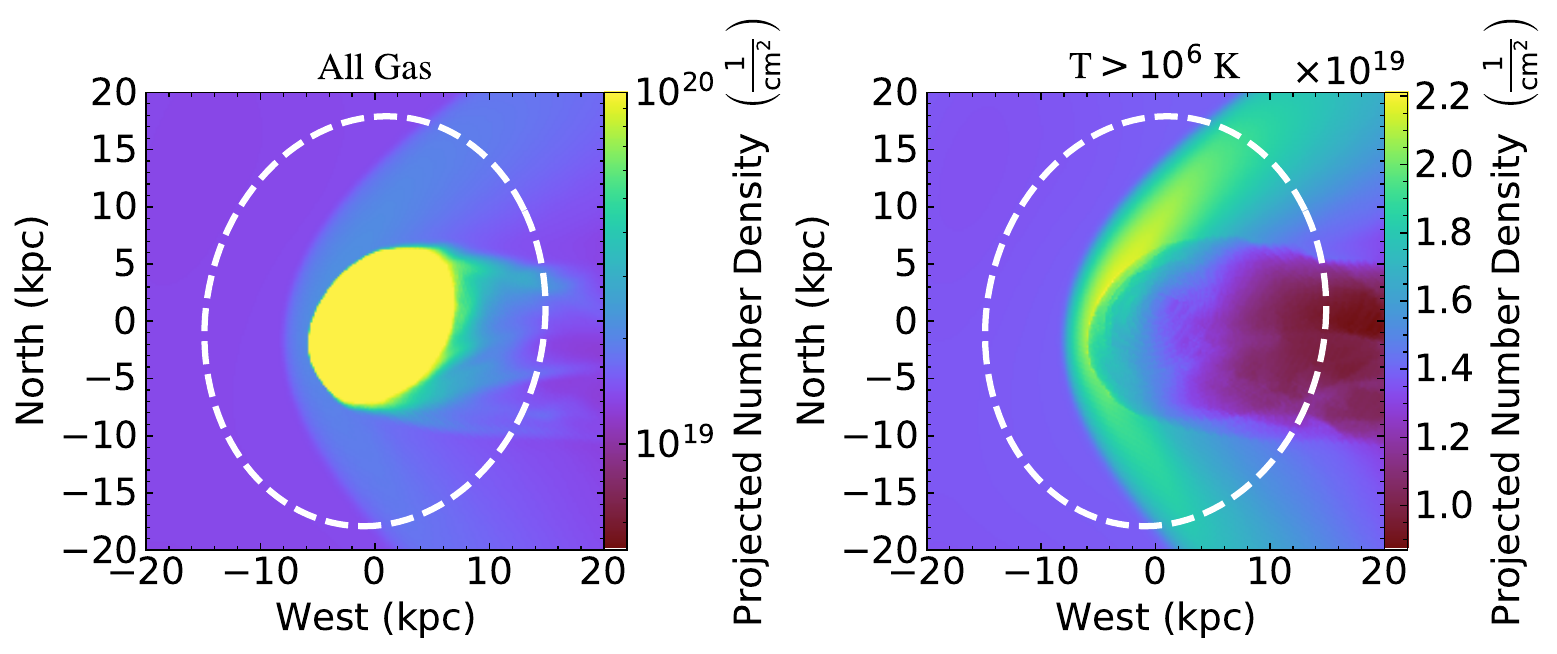}
\caption{Number density of all gas (CGM and LMC disk gas; left) and only hot CGM gas (T$>10^6$ K, right) projected along our line of sight towards the LMC, shown on log and linear scales, respectively, to highlight the dynamic range of density. The cold gas of the disk dominates the total density. In the hot gas, a bow shock induced by the galaxy's supersonic motion accounts for most of the differences in density. This shock subtends 10s of kiloparsecs in the simulation, and the column density is $\sim2$ times larger in the shocked CGM than it is in the unshocked regions. In both panels, the observed extent of the LMC stellar disk ($r\sim$18 kpc, see \citealt{Mackey2018}) is shown as a dashed white line, illustrating that both the gas disk and the shock front are located at significantly smaller radii than the stellar disk. The stand-off radius for the shock is $R_{so}=6.7$ kpc, while the gas disk is truncated at $\sim$6.2 kpc \citep{Salem2015}.
\label{fig:gas_los}}
\end{figure*}

\section{Description of the Simulation} \label{sec:sim}

To place constraints on the shape and structure of the LMC's bow shock, we utilize the same {\it Enzo} \citep{Bryan2014} hydrodynamic simulation as in \cite{Salem2015}. This simulation was performed to compare the effects of ram pressure stripping on the truncation of the LMC's gas disk to observations, thereby constraining the properties of the CGM at the location of the LMC. Their fiducial simulation, which best matches the disk truncation, is also very well suited to studying other effects of the LMC's supersonic motion through the CGM such as the aforementioned bow shock. While the full details of the simulation are described in detail in \cite{Salem2015}, here we summarize the key simulation conditions and highlight any differences in our implementation.

The simulation uses a ``wind tunnel" approach by placing the LMC at rest in a box and modeling the MW CGM as a headwind with evolving velocity and density structure meant to mirror the actual CGM conditions the LMC passes through along its orbit. The orbit of the LMC is a first infall scenario, such that the LMC has not made a previous orbit about the MW as in \citet{Kallivayalil2013}.

In the original simulation, the LMC was placed at the coordinates (20, 20, 20) kpc in a $60\times60\times60$ kpc$^3$ box. In our implementation, we re-ran the simulation using a $100\times100\times100$ kpc$^3$ box, placing the LMC at the same relative location. The larger box size was chosen to prevent the bow shock from reaching the edges of the simulation box and interacting with boundary conditions in the present day simulation snapshot. The resolution of the simulation is consequently lower by a factor of 1.6 compared to \cite{Salem2015}, but as we are concerned with the response of the larger scale CGM, the lower resolution does not impact our findings. 

The simulation includes self-gravitating gas in the form of the LMC gas disk and the ambient MW CGM. The simulation does initialize with a small ($5\times10^6 \ M_\odot$) LMC CGM, but that halo is quickly swept away by the wind and is thus not assumed to survive the LMC infall. The simulation does not utilize radiative cooling or star formation feedback. The gravitational potentials of the LMC stars and dark matter are treated as static potentials. The LMC dark matter halo is modeled using a spherical profile with $\rho_0=3.4 \times 10^{-24} \ \mathrm{g \ cm^{-3}}$, $r_0=3.4$ kpc, and $M\mathrm{(100 kpc)} = 5 \times 10^{10} M_\odot$ following \cite{Burkert1995}. The stellar component is modeled as a Plummer-Kuzmin Disk \citep{Miyamoto1975} with $M_\star=2.9 \times 10^9 M_\odot$, $a_\star=1.7$ kpc, and $b_\star=0.34$ kpc. The initial gas distribution is set as an exponential disk with $M_\mathrm{gas}=5 \times 10^8 M_\odot$, $a_\mathrm{gas}=1.7$ kpc and $b_\mathrm{gas}=0.34$ kpc (following the stellar distribution), and $M_\mathrm{tot}=7.2 \times 10^8 M_\odot$ \citep{Tonnesen2009}. See Table 1 in \cite{Salem2015} for more details.

\begin{figure*}
\includegraphics[width=\textwidth]{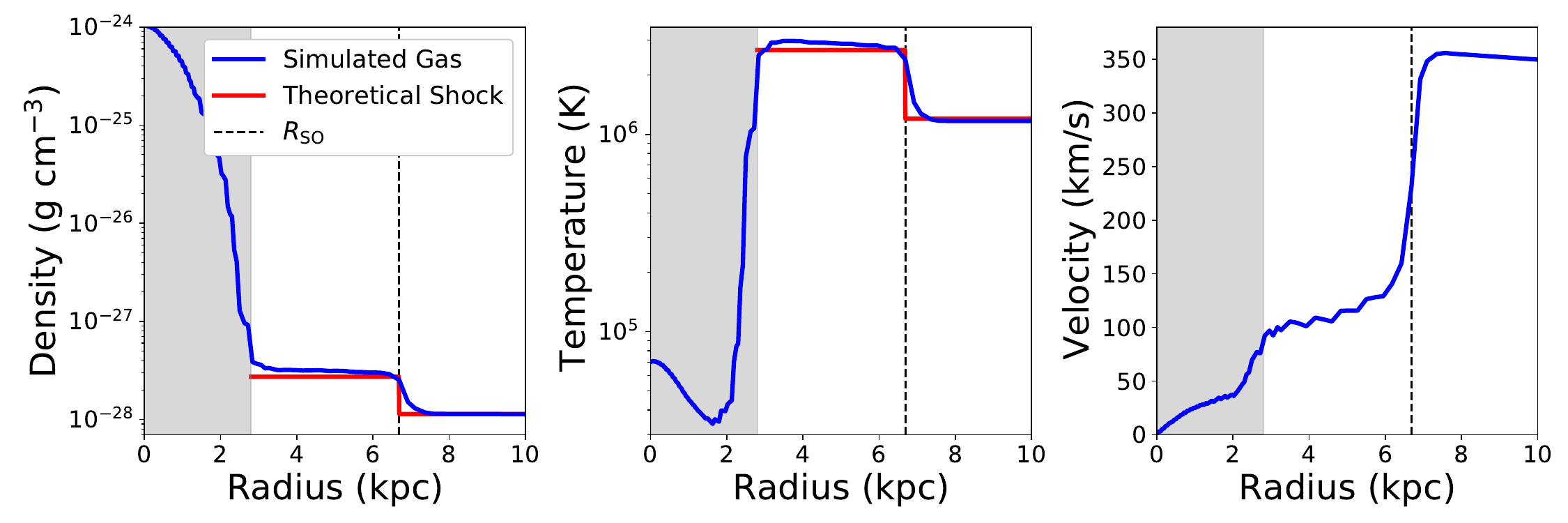}
\caption{Radial profiles (blue) for the density (left), temperature (center), and velocity (right) along a vector that points from the LMC disk center in the direction of the LMC's velocity vector. The red line shows the expected jump values for the LMC's density and temperature, assuming a CGM temperature of 1.19$\times 10^6$ K and a velocity of 350 km/s as in the simulation. We additionally show the stand-off radius, $R_\mathrm{SO}$, as calculated by the algorithm outlined in Section \ref{subsec:shape}. The theoretical predictions match the simulated shock conditions well in density and temperature. Additionally, there is a clear discontinuity in the velocity structure of the gas in the shocked region wher the shocked gas is moving at $\sim$100 km/s relative to the LMC, significantly slower than the ambient CGM. The grey shaded region shows the extent of the LMC gas disk.
\label{fig:shock_conditions}}
\end{figure*}

\section{The Bow Shock of the LMC}

Following \cite{Salem2015}, in the simulation, gas is accelerated towards the LMC, and the gas just ahead of the LMC at present day is in fact moving at a speed of $\sim$350 km/s relative to the LMC center of mass, slightly higher than the value of 320 km/s reported in \cite{Kallivayalil2013}. In order to be consistent with the simulation, we utilize this value in conjunction with the CGM conditions to calculate the Mach Number ($M$):

\begin{equation}
    M = \frac{\lvert \vec{v} \rvert}{c_s} = \lvert \vec{v} \rvert \sqrt{\frac{m}{\gamma k T}}
\end{equation}

\noindent where $\gamma=5/3$, $m$ is the effective mass of hydrogen ($\mu \times 1.67 \times 10^{-27}$ kg where $\mu=0.6$), and $k$ is the Boltzmann constant $1.381 \times 10^{-23} \mathrm{\frac{J}{K}}$. In the fiducial model for the CGM in \cite{Salem2015}, the gas temperature of the CGM is 1.19$\times 10^6$ K. At this temperature the sound speed is 165 km/s, assuming $\gamma=5/3$; consequently the LMC in the simulation is moving at Mach $\sim2.1$. This temperature assumption is well supported by x-ray observations of O {\footnotesize VII} and O {\footnotesize VIII} that can be well modeled in a hot CGM plasma \citep{Miller2013, Miller2015}. We note that because the Mach number scales as $T^{-1/2}$, if the CGM is colder than we assume while still being thermally supported, we would actually predict a stronger shock by as much as an order of magnitude that would keep the shocked gas at $T\gtrsim10^6$ K. However, this argument would not hold in the case of a cooler cosmic ray dominated CGM \citep[e.g.,][]{Ji2020}, where the pressure support would not be primarily thermal. Given that there is strong observational evidence for CGM temperatures that would result in the supersonic motion of the LMC, we predict that a bow shock will lead the system. Throughout the rest of this letter, we quantify the strength and shape of the shock assuming the fiducial CGM conditions from \cite{Salem2015}.

In Figure \ref{fig:gas_los}, we show the simulated gas density of the LMC and CGM projected along the line of sight. The plots show all gas (LMC disk gas and CGM; left) and only the hot CGM gas ($>10^6$ K; right). See Table 3 in \cite{Salem2015} for the detailed coordinate transformations that takes the simulated LMC from the box frame to a line-of-sight frame, where the LMC velocity vector is aligned with the 3D Galactocentric velocity vector and the disk is inclined correctly in Cartesian coordinates from our viewing perspective at the location of the Sun.   

It is clear that shocked (higher density and temperature) CGM gas surrounds the simulated LMC.  %subtends the simulated motion of the LMC. 
The projected column of hot gas is $\sim2$ times larger than that of the column through undisturbed ambient CGM. Additionally, while the shock exists in front of the LMC gas disk, it is predominantly located inside the LMC's observed stellar disk, which is represented as a white dashed circle of radius 18.5 kpc \citep{Mackey2018}. 

In the following sections, we use this simulation to explore the influence of the shock on the CGM gas as the LMC falls into the Milky Way's potential.

\begin{figure*}
\centering
\includegraphics[width=\textwidth]{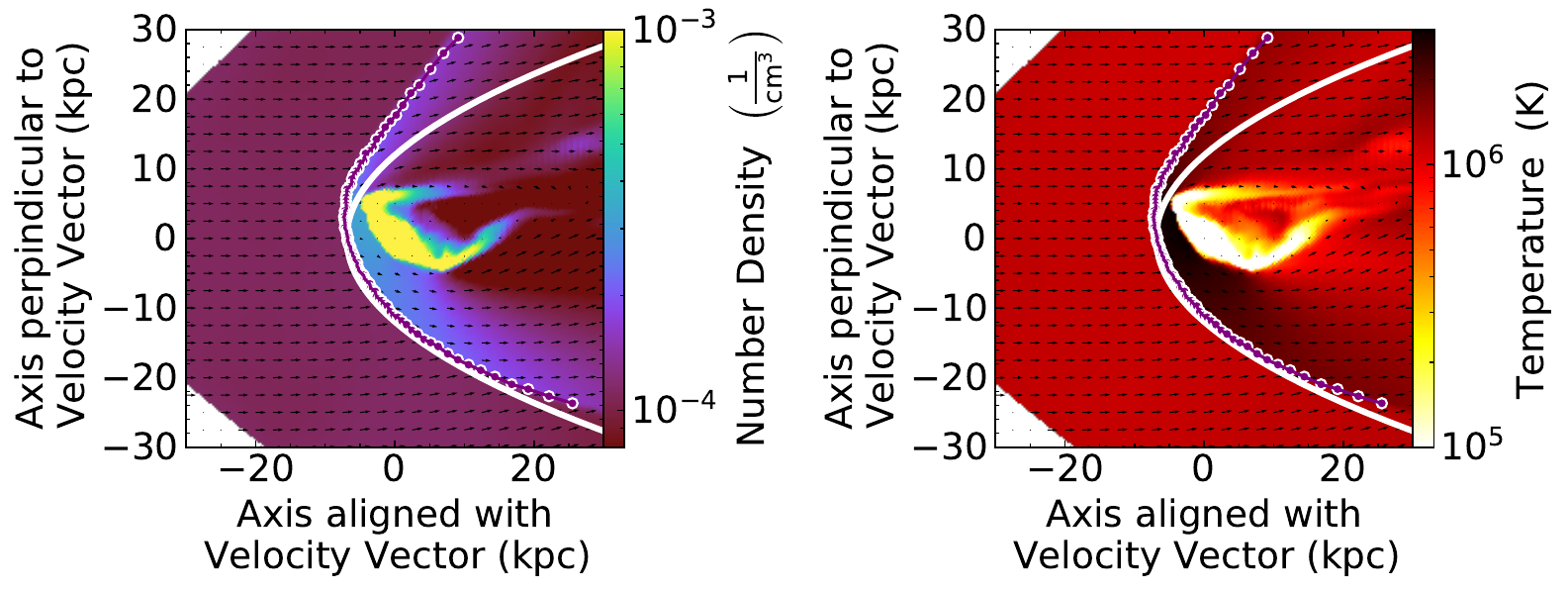}
\caption{The density (left) and temperature (right) in a slice through a single resolution element of the simulation volume ($\sim0.5$ kpc), which is oriented parallel to the velocity vector along the x-axis. The color scale is chosen such that the density and temperature are saturated for the LMC disk to highlight the shock structure. The grid of vectors indicate the velocity of the CGM relative to the LMC. As expected, the magnitude of the velocity at the shock drops near the boundary, indicating that the shock is traveling with the LMC (see also Figure \ref{fig:shock_conditions}). The purple points show the shock locations found in the simulation by our shock finding algorithm. The white line shows the expected shape for a bow shock around a spherically symmetric object \citep[e.g., a star, see][]{MacLow1991,Cox2012} with the same stand off radius as the LMC ($R_{so}=6.7$ kpc). The shapes are similar, but the orientation of the LMC disk relative to its velocity vector results in an asymmetric shock.
\label{fig:vel_slice}}
\end{figure*}

\section{Shock Conditions and Shape} \label{sec:shape}

\subsection{Shock Conditions}

Due to the complicated nature of hydrodynamic equations, complex behavior such as the formation of a bow shock around the inclined LMC disk cannot be modeled analytically. However, the leading edge of a shock can be approximated as one-dimensional, and therefore can be analyzed to first order using fluid equations. The Rankine-Hugoniot jump conditions characterize the strength of the jump in terms of temperature ($T$) and density ($\rho$) ratios based on the Mach number ($M$), and the heat capacity ratio ($\gamma$).

\begin{equation}
    \frac{T_2}{T_1}=\frac{[(\gamma+1)+2\gamma(M_1^2-1)][(\gamma+1)+(\gamma-1)(M_1^2-1)]}{(\gamma+1)^2M_1^2} \label{equation:tjump}
\end{equation}

\begin{equation} \label{eqa:dens_jump}
    \frac{\rho_2}{\rho_1}=\frac{(\gamma+1)M_1^2}{(\gamma+1)+(\gamma-1)(M_1^2-1)}
\end{equation}

To compare the conditions in the simulation to the theoretical jump conditions, we extract the density, temperature, and pressures of the simulated gas along a ray which points from the center of the LMC along the direction of the LMC velocity in the present day simulation slice. In Figure \ref{fig:shock_conditions}, we show the density, temperature, and pressure along this ray. In red, we show the predicted shock strength from the Rankine-Hugoniot conditions assuming $\gamma=5/3$, a CGM temperature of 1.19$\times 10^6$ K (see Figure \ref{fig:shock_conditions}), and a velocity of 350 km/s (corresponding to a Mach number of $\sim2.1$), the present day effective velocity of the LMC in the simulation. The density, temperature, and pressure jumps in the simulation match the predicted ones well, demonstrating that the resultant shock in the simulation is consistent with theoretical predictions. We note that while the simulation does not utilize radiative cooling, we do not expect this to affect the shock as the cooling time for the gas is on the order of a 4-40 Gyr (assuming the ambient CGM temperature and density in the fiducial simulation and a cooling rate of order $\sim1-10 \times 10^{-23} \ \mathrm{erg \ s^{-1} \ cm^3} $, spanning a wide range of possible metallicities of the gas), much longer than any other relevant timescales within the simulation. We expect this to be true for the shocked gas even in the case that the ambient CGM is significantly cooler than we assume in the simulation, as the higher Mach number in conjunction with Equation \ref{equation:tjump} implies that the temperature of the shocked region should be $\geq 10^{6}$ K regardless of the temperature of the ambient CGM.

Additionally, we measure the stand-off radius of the shock front, the distance from the center of the LMC to the shock front along the direction of the LMC velocity vector, and compare to empirically motivated lab measurements of shock shape. \cite{Billig1967} showed that the stand-off radius for a shock generated around a rigid body, $R_{so}$, takes the form $R_{so}/R = 1 + 0.143e^{3.24/M}$ where $M$ is the Mach number. While the LMC disk is neither a rigid body nor is it symmetric, we can approximate its radius as $R_\mathrm{eff} = R_\mathrm{gas} \mathrm{sin}\theta$ where $R_\mathrm{gas}$ is $\sim6$ kpc and $\theta$ is the angle between the LMC's angular momentum vector and its velocity. Substituting these values along with the Mach number, 2.1, we obtain an estimate for the stand off radius of $\approx5.75$ kpc. This is within $\sim20\%$ of the stand-off radius in the simulation, 6.7 kpc (see Figure \ref{fig:shock_conditions}). Throughout the rest of this work, we adopt the empirically derived 6.7 kpc as our fiducial bow shock stand-off radius at present day. 

\begin{figure*}
\includegraphics[width=\textwidth]{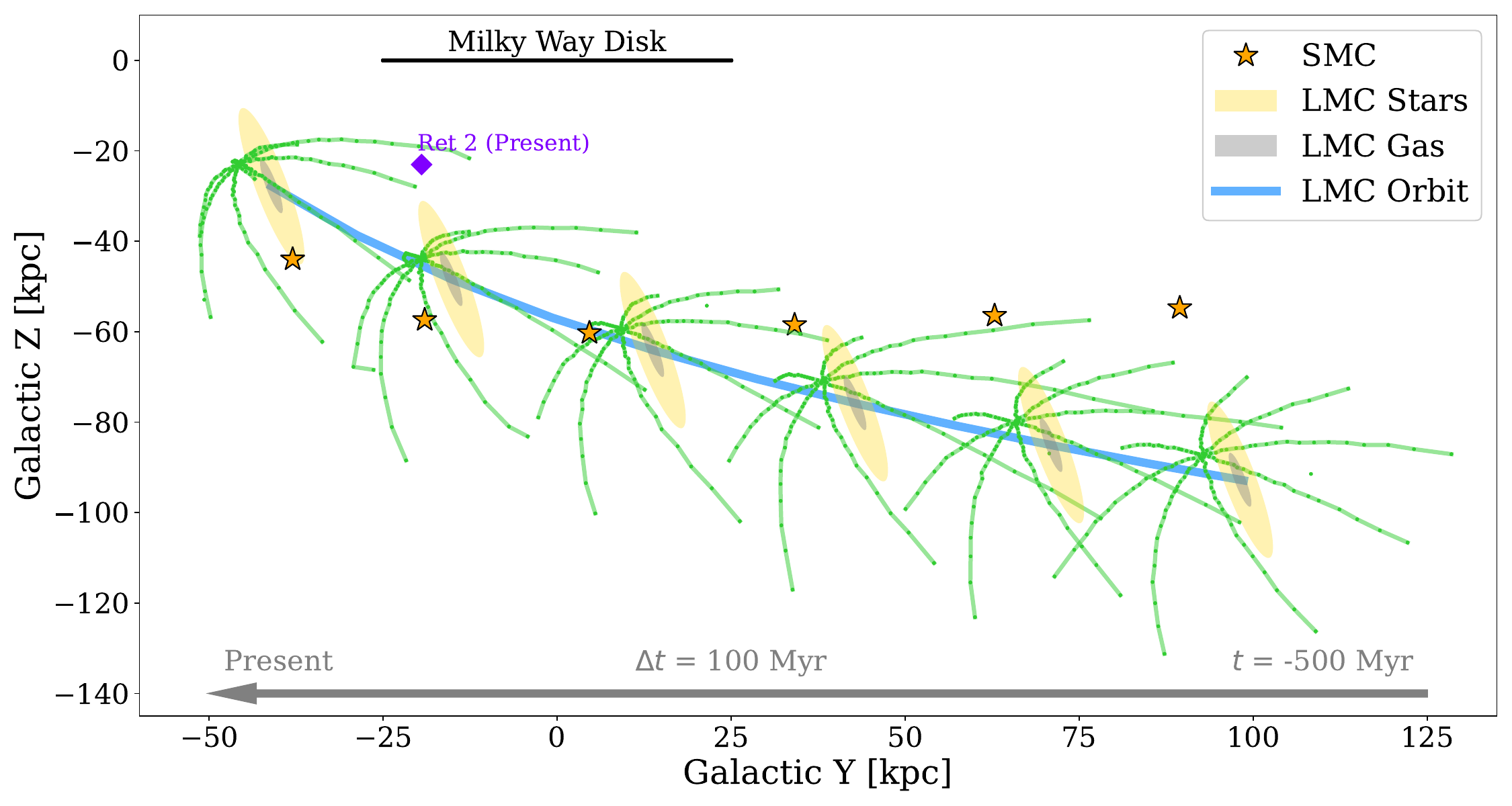}
\caption{The orbital path of the LMC (gas: grey, stars: yellow) and its evolving bow shock over the last 500 Myr in the YZ Galactocentric plane. Because the simulations allow for time evolving conditions of the wind density and speed that mimics the LMC's orbit, we can utilize our shock finding algorithm to identify the location of the shock along the past orbit of the LMC. The extended shock (green points) has carved out a significant volume of the CGM in the southern sky over this period of time; we estimate that as much as $\sim8\%$ of the gas in the southern hemisphere may have interacted with the boundary of the shock during this time period. Additionally, we show the location of the SMC at each timestep (orange star) and the position of Ret 2 at present (blue diamond).
\label{fig:shock_sat}}
\end{figure*}

\subsection{Shock Shape} \label{subsec:shape}

While the empirical shapes of shocks surrounding rigid, symmetrical bodies are well characterized \citep[e.g.][]{MacLow1991, Cox2012}, the diffuse nature of the LMC gas as well as the angle between the disk orientation and infall velocity results in an asymmetric shock shape that cannot be modeled analytically. In Figure \ref{fig:vel_slice}, we show slices ($\sim0.5$ kpc, one resolution element thick) in density (left) and temperature (right) centered on the LMC in the present day simulation,. The x-axis in both slices is aligned with the bulk flow simulation velocity (shown as black arrows), which is the inverse of the LMC velocity at this time. The shock boundary can be clearly seen in both density and temperature. However, because of the inclination of the LMC disk relative to its motion, the shock is not symmetric. This can be clearly seen in the comparison to the analytic solution for a shock around a rigid body \citep{MacLow1991, Cox2012}:

\begin{equation}
    y = \sqrt{3}R_{so}\sqrt{1-\frac{x}{R_{so}}},
\end{equation}

\noindent where $R_{so}$ is the stand-off radius, or the radius of the shock along the velocity vector of the object's supersonic motion, x is the coordinate aligned with the velocity of the wind, and y is a vector perpendicular to x. 

In Figure \ref{fig:vel_slice} we show this empirical shock shape for $R_{so}=6.7$ kpc (white line). %, which we measure from the simulation, and we center our coordinate system at the LMC's center. 
In the regions directly preceding and below the LMC in this slice, the analytic model matches the shape of the shock quite well. However, the disk inclination causes the simulated shock to have a wider opening angle in the regions nearest to the disk, resulting in a strongly asymmetric simulated shock. %and the predicted shock in the simulation is strongly asymmetric.

In order to empirically measure the shock's shape in the simulation, we automate the finding of the shock by drawing rays in four planes parallel to the LMC's velocity from the center of the LMC to the edge of the box at angles in the plane spanning $-140^\circ$ to $140^\circ$ where $0^\circ$ is defined as the direction of the LMC's velocity in the slice. Along each ray, we define the shock location as the first point from the end of the ray where the temperature reaches 25\% of the predicted jump in temperature from the Rankine-Hugoniot conditions (see Equation \ref{equation:tjump}), using the velocity of the LMC and the temperature of the gas to derive the Mach number at that time. This procedure finds the location of the shock along a line of sight with a high degree of accuracy, although it can sometimes fail in the early times of the simulation when a line of sight moves through stripped LMC gas behind the LMC. The results of this automated shock finding on a present day simulation slice are also shown in Figure \ref{fig:vel_slice} as purple points, which, in contrast with the shape prescribed for a symmetric shock around a sphere, trace the edge of the shock extremely well. 

In addition to measuring the shape of the shock in the present day slice, we can also use the simulation to measure the evolution of the shock as a function of time. In Figure \ref{fig:shock_sat} we show the location of the LMC+shock system in six slices ranging from present day (left) to 500 Myr ago (right), assuming that the orientation of the disk remains constant throughout the infall for ease of visualization (though we note that the orientation of the disk is not expected to change significantly via LMC/SMC interactions during the infall, see \citealt{Besla2012}). As the velocity of the LMC increases as it approaches pericenter, the shock's opening angle becomes smaller and the stand-off radius becomes smaller, but even 500 Myr ago the supersonic motion of the LMC clearly results in a shock that subtends the extent of the LMC. Note that there are a few measurements of the shock location that are not contiguous with the otherwise continuous shocks. These tend to result from failures in our shock-finding algorithm as our ray moves through trailing gas stripped material from the LMC disk, especially in early simulation slices.

The physical extent of the shock is significantly larger than the LMC gas disk.  The shock extends $\sim30$ kpc from the center of the LMC in the plane traverse to the LMC velocity, in contrast with the 6 kpc gas disk. Additionally, the leading edge of the shock is located well within the LMC's stellar disk ($\sim$18 kpc, see \citealt{Mackey2018}) due to the truncation of the gas from ram pressure during infall \citep{Salem2015}. We predict that along the past orbit of the LMC, the MW CGM should be $\sim2-4$ times as dense as the ambient CGM at that radius as evidence of the shock's passage.

\section{Observability of the Shock} \label{sec:observability}

Our simulation predicts that ram pressure stripping and the bow shock will result in sharp boundaries between the LMC gas disk, the shocked region, and the ambient CGM. As such, ionized gas probes such as $H\alpha$ emission, which are strongly sensitive to the density and temperature of the gas, may trace the different physical conditions in these regions. While it is likely that $H\alpha$ emission in the star forming regions of cold ISM within LMC will dominate the total flux in the region and the shock itself is too hot to be a significant source of $H\alpha$ flux, we propose that interactions between the shocked CGM gas and cold clouds surrounding the LMC may ionize hydrogen and result in an extended $H\alpha$ signature that that extends beyond the cold neutral gas and is systematically moving at the same velocity as the LMC.

\begin{figure}
    \centering
    \includegraphics[width=0.45\textwidth]{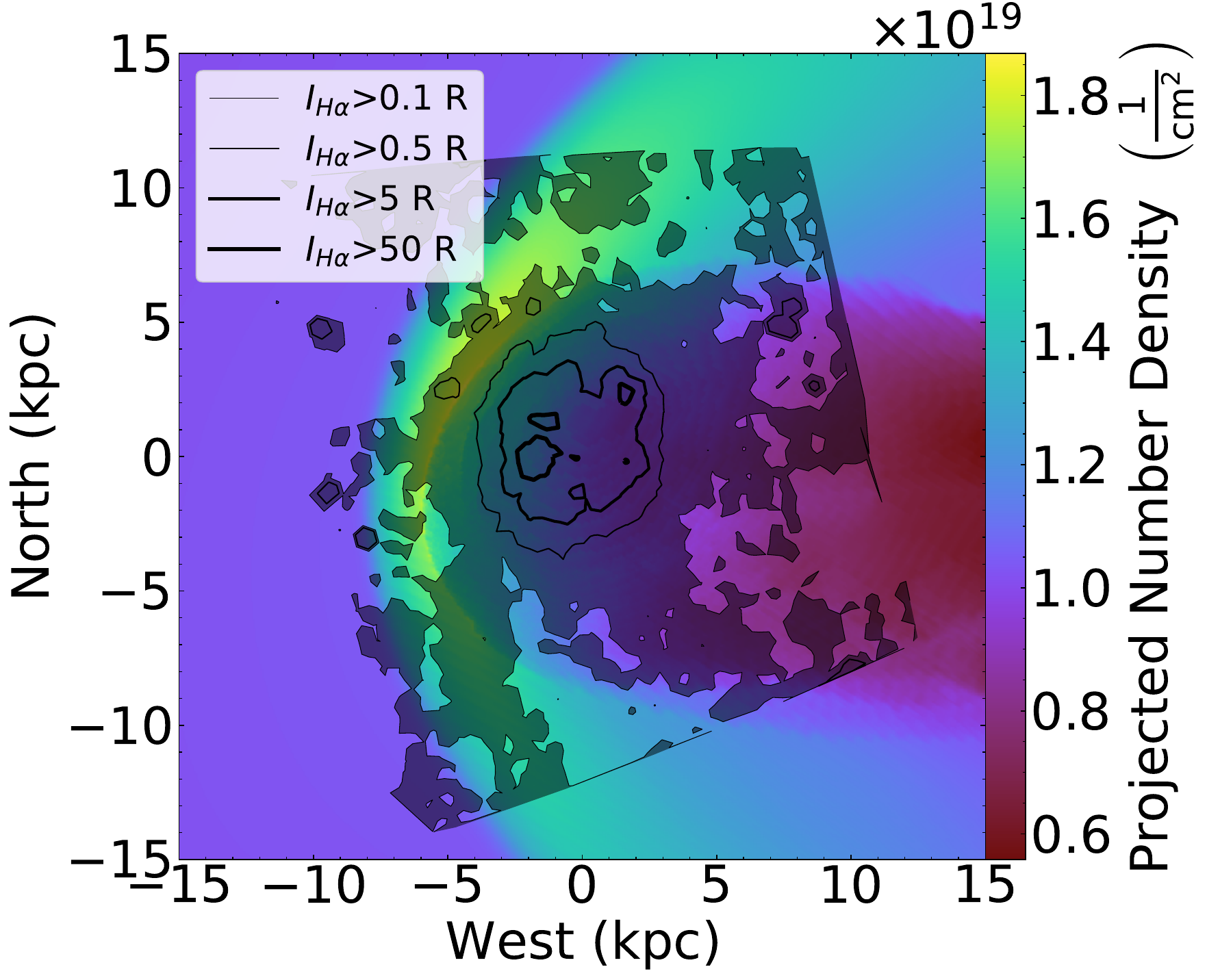}
    \caption{The simulated LMC-shock system presented in this work (as in Figure \ref{fig:gas_los} for $T>10^6$ K) shown with observed $H\alpha$ contours measured with the Wisconsin $H\alpha$ mapper \citep{Smart2023}. We show integrated $H\alpha$ contours encompassing 0.1, 0.5, 5, and 50 Rayeleigh intensity integrated between -130 km/s and 110 km/s in the LMC frame of reference. We suggest that the diffuse $H\alpha$ structure leading the LMC disk may be associated with the shock, providing a clear explanation for the $H\alpha$ structure that is more extended and follows a different morphology than the neutral HI gas.}
    \label{fig:halpha}
\end{figure}

Recent measurements of the LMC using the Wisconsin $H\alpha$ Mapper (WHAM) show exactly such a feature. \cite{Smart2023} measured spatially extended $H\alpha$ along the direction of the LMC's leading edge at the systematic velocity of the LMC (see Figure 7 in that work). In Figure \ref{fig:halpha}, we overlay an H$\alpha$ intensity map of the LMC obtained from the WHAM survey \citep{haffner03} that was studied in \cite{Smart2023} on our projected density map of hot ($T>10^6$ K) gas. To do so, we follow the procedure in \cite{Smart2023} to integrate the WHAM data from $-130$ to $110$ km s$^{-1}$ in the velocity reference frame of the LMC (see their Equation 5). We then project the H$\alpha$ map into an orthographic projection based on Equation 1 in \cite{gaia21} \citep[see also][]{choi22} assuming an LMC distance of 50.1 kpc \citep{freedman01} and center of mass at (RA, DEC)=(78.76$^{\rm o}$, $-69.19^{\rm o}$) \citep{vanderMarel14}, which were the same values used in \cite{Salem2015}. We show contours for intensities of 0.1, 0.5, 5, and 50 Rayleighs with proportionally increasing line widths. We find that the asymmetric feature along the leading arm is in excellent agreement with the predicted shock front. While the majority of the ionized gas is located within the gas disk of the LMC, the $I_{H\alpha}>0.1$ R contours of the ionized hydrogen clearly extend into the shocked region past the extent of the neutral gas disk and truncate near the edge of the shock. While the analysis presented here is purely morphological (as the exact extent of the shock is sensitive to the temperature of the ambient CGM, which is not well constrained in \citealt{Salem2015} and the distribution of cold clumps in the region surrounding the LMC is a significant unknown), the alignment of this asymmetric $H\alpha$ with our predicted shock provides a tantalizing piece of observational evidence that supports the shock's existence. A more detailed prediction from this simulation of the $H\alpha$ brightness within a shocked medium containing cold molecular clouds is currently underway.

Additionally, our simulations allow us to roughly predict the column density excess caused by the presence of the shock, assuming an otherwise isotropic and constant density CGM. To do so, we generate two gas density profiles, one along a line of sight through our simulation box that points directly through the shocked medium at the stand off radius ($\sim$ 6 kpc along the LMC's velocity vector), and another 15 kpc in front of the LMC, well past the stand off radius of the shock. Computing the column density over a distance of 60 kpc, we find that the shock-tracing and non-shock tracing column densities are $\sim3\times10^{19} \ \mathrm{cm}^2$ and $\sim2\times10^{19} \ \mathrm{cm}^2$, respectively. As such, we expect that the enhancement in column density along a line of sight that intersects the leading edge of the shock should be $\sim 10^{19} \ \mathrm{cm^2}$ for the $T\sim10^6$ K assumed in our simulation. As we show in Figure \ref{fig:gas_los}, this enhancement in column density will not be restricted to the region directly in front of the shock, but would be observable on scales of $\sim30$ kpc tracing out the shape of the shock.

Note, we have estimated the non-thermal emission to be produced by the LMC bow shock via Fermi processes. To estimate the non-thermal emission produced via Fermi process in the bow shock of the LMC, we calculated synchrotron radiation ($\lesssim10^{-12} \ \mathrm{erg/s/cm^2/sr}$), emission from inverse Compton scattering ($\lesssim10^{-12} \ \mathrm{erg/s/cm^2/sr}$), and synchrotron self-Compton emission ($\lesssim 10^{-14} \ \mathrm{erg/s/cm^2/sr}$) following the methodology outlined by \cite{Wang2015}.  However, we find that the non-thermal emission produced by the LMC shock will not be strongly observable at any wavelength.

\section{Discussion} \label{sec:discussion}

\subsection{Satellite Bow Shocks as Sources of Mixing and Heating within the CGM}

The infall of the LMC has been shown to exert significant perturbations on the density and kinematics of the MW's dark matter and stellar halos \citep{Garavito2019,Conroy2021}, highlighting that massive satellite galaxies can give rise to non-equilibrium halo structures. We posit that the LMC bow shock may similarly play a role in the structure of CGM gas, as the bow shock associated with the LMC can sweep out a significant volume as it falls into the MW. 

In Figure \ref{fig:shock_sat}, we show the physical extent of the LMC/shock system at six time steps in relation to the static Milky Way disk, illustrating the path of the shock over the past 500 Myr. To calculate the volume of gas that may have been shocked during the LMC infall, we integrate along the LMC path, approximating the shock as an $r=30$ kpc disk that is oriented perpendicular to the path of the LMC at all points in the orbit. Over the past 500 Myr, the LMC has traveled $\sim160$ kpc, meaning that $\sim5.5 \times 10^5 \ \mathrm{kpc^3}$ of CGM gas will have interacted with the bow shock and potentially been perturbed during this period. This represents $\sim8.2\%$ of the total CGM volume bounded within the r$\sim140$ kpc hemisphere of the southern sky that the LMC has occupied over the past 500 Myr. 

While the jump in density in temperature induced by the shock is modest given the CGM conditions in the simulation, we posit that it still may influence the CGM dynamics in two key ways. Firstly, the shock interaction with small scale fluctuations in the CGM may provide an important source of turbulent pressure support to the CGM at large radii. This effect has been observed in cosmological simulations \citep[see Figure 12 in][]{Lochhaas2023}, and we posit that it is likely occurring in our own backyard on significant physical scales. Secondly, if there are existing cool gas structures that the shock sweeps past, the post-shocked gas will drive a velocity shear between the hot post-shocked gas and the pre-existing cool gas, which could alter cloud survival conditions \citep[e.g.,][]{Gronke2018, Abruzzo2023} such that small cool clouds that could survive in the ambient CGM would be destroyed in the interaction with the shock. We note that the volume carved out by the interaction will grow strongly with time, as the LMC is only on its first infall \citep{Besla2012} and the shock will continue to sweep out CGM gas until the galaxies merge.

In other words, the infall of a massive satellite may be a significant driver of the mixing of dense clouds with more diffuse CGM gas in the CGM, particularly if the satellite has made multiple orbits (e.g. the case of the Sagittarius dwarf about the MW and the progenitor of the Giant Southern Stream about M31). This mechanism for mixing may be even more significant in cluster environments, where both gas densities and galaxy speeds are higher.  

\subsection{Implications for the origin of ionized gas associated with the Magellanic System}

The Magellanic Clouds are currently contributing $4.87 \times 10^8$ M$_\odot$ (d/55 kpc$)^2$ \citep{Bruens2005} of neutral HI to the MW's CGM, where d is the assumed Galactocentric distance to the gas. This gas is being removed from the Clouds and forms large-scale gaseous structures, like the Magellanic Bridge and Stream \citep{Putman2003, Nidever2010}. But this HI gas is only $\sim 0.5-2\%$ of the MW's total CGM mass budget \citep[3-10 $\times 10^{10}$ M$_\odot$;][]{Faerman2022}, assuming a distance to the Magellanic Stream of d=55 kpc. 

There is, however, also a significant ionized gas component associated spatially and kinematically with the Magellanic Stream \cite[$\sim 2 \times 10^9$ M$_\odot$ (d/55 kpc)$^2$;][]{Fox2014}, Magellanic Bridge \citep[0.7-1.7 $\times 10^8$ M$_\odot$;][]{Barger2013}, and surrounding the LMC \cite[0.6-1.8 $\times 10^9$ M$_\odot$;][]{Smart2023}. It has been posited that much of this gas originates in a ``Magellanic Corona" \citep{Krishnarao2022} that is being removed through a combination of ram pressure and tidal stripping \citep{Lucchini2020}. 

% However, as the LMC circumgalactic medium should long since have dissipated due to ram pressure stripping \citep{Salem2015}, it is unlikely that the Magellanic Corona was brought into the MW CGM with the LMC/SMC system.

The ionized gas component surrounding the neutral Magellanic Stream is a large mass budget that could contribute between 3-10\% of the MW's total CGM mass, assuming an average d=55 kpc to the Stream. If the Stream is located on average at a larger distance of $\sim$100 kpc \citep{Besla2012}, the associated ionized gas could contribute 8-26\% of the total CGM mass budget. This is a non-negligible contribution to the CGM, making it important to understand the origin of this ionized gas.

In this study, we have illustrated that MW CGM gas ahead of the LMC should be accelerated to the LMC's systemic speed and form a bow shock. We posit that the interactions between this hot, shocked gas and cooler clouds within the CGM could result in elevated ionization in the region trailing the LMC infall. This means that the bow shock could explain a significant portion of the ionized gas observed surrounding the LMC, without appealing to an existing LMC CGM that resists dispersal by ram pressure (Section \ref{sec:observability}), providing an alternative explanation for the ionized gas observed at high ionization states associated with a ``Magellanic Corona". Indeed, the predicted enhancement in column density along the the lines of sight through the shock (see Figure \ref{fig:gas_los} and Section \ref{sec:observability}) are within $\sim$ a factor of 2-3 of the inferred coronal columns from Figure 3 in \cite{Krishnarao2022}, indicating that a bow shock provides a plausible explanation for a significant amount of the excess material observed around the LMC.

This scenario further implies that, as the LMC moves through the MW's CGM, it would leave a trail of ionized/shocked CGM gas (see Figure \ref{fig:shock_sat}). In other words, the observation of ionized gas spatially trailing the past orbit of the Clouds and moving at speeds consistent with the neutral Stream {\it does not require} that the ionized gas be removed from the Clouds themselves. Rather, this ionized gas could plausibly result from the interface between the MW's CGM and the LMC's gas disk, or the neutral hydrogen Stream, as both structures move through the MW's CGM. The observed line ratios of CIV, OVI, SiIV, etc, surrounding the Stream \citep{Fox2010, Fox2014} strongly point to interface/mixing layers between the neutral HI in the Stream and the MW CGM. However, shocks can produce similar ratios \citep[e.g.,][]{Wakker2012}, meaning that it is plausible that some of the observed ionized gas stems directly from the interaction between the bow shock and the MW CGM.  

This idea is important for understanding the total mass budget of the Magellanic Clouds at infall, which has consequences for our understanding of the baryon fraction in low mass galaxies. Many models of the Magellanic Stream require the bulk of its gas mass to originate from the SMC \citep[e.g.,][]{Besla2012}. If we assume that 50\% of the ionized + neutral gas budget of the Magellanic Stream and Bridge originate from the SMC, and that the SMC had an infall halo mass of order $2 \times 10^{10}$ M$_\odot$ \citep{Besla2010}, and current neutral gas mass of $4 \times 10^{8}$ M$_\odot$ \citep{Bruens2005}, the SMC would need to have a baryon fraction of $\sim$8-23\% at infall, depending on the distance to the Magellanic Stream (55-100 kpc). Such baryon fractions are much higher than the 3-5\% in MW-like galaxies. In the shallower halo potentials of low mass galaxies, stellar winds should be more efficient, making baryon fractions even lower, not higher \citep{Besla2015}. Importantly, the total ionized gas mass associated with the Stream has been used as an argument against tidal-driven models for the origin of the Stream \cite[see review in ][]{donghia2016}, but if much of that gas is of MW CGM origin, the mass discrepancies between the models and observations are significantly less of an issue.  

This argument about the origin of the Magellanic corona does not imply that the LMC did not once possess an ionized CGM. Indeed, cosmological simulations of isolated LMC mass analogs do possess CGMs \citep{Jahn2022}. Thus, some component of the present-day ionized gas surrounding the Magellanic Stream could come from the LMC's CGM that has been stripped. This is the case in the simulations of \cite{Salem2015}, who initialize their LMC with a low mass CGM ($5\times 10^6$ M$_\odot$) that is almost immediately swept into an ionized tail owing to ram pressure. Given that the truncated HI disk of the LMC provides clear evidence of direct impact from ram pressure \citep{Salem2015}, our simulation suggests that the ionized gas at high ionization states within 20-30 kpc of the LMC observed by \cite{Krishnarao2022} is unlikely to be the primordial CGM gas that the LMC possessed at infall.

We note that the initialized LMC halo mass in our simulations is $\sim3$ orders of magnitude lower than that in recent simulations where the LMC CGM has survived the infall into the LMC halo as a Magellanic Corona \cite{Lucchini2020}. In these simulations, some of the initial LMC CGM does survive the interactions with the MW CGM. However, our simulations are designed to match the observed truncation of the LMC's gaseous disk \cite{Salem2015}. Ram pressure stripping is dependent on the gas surface density of the infalling satellite. In order to achieve a truncation radius for the HI disk that is compatible with observations, the ram pressure faced by the LMC must be sufficiently strong to remove high gas density material in the disk at distances of $\sim$5-6 kpc. By definition this ram pressure must be strong enough to also remove the initial CGM gas, which is always lower density than the gas disk itself. Indeed in \cite{Salem2015} the simulated low density LMC CGM is removed by ram pressure before the gas disk is truncated. This would be true regardless of the LMC halo mass, although the removed CGM may still be retained at large distances in a more massive LMC model. As a result, we propose here instead that the ionized gas within 20-30 kpc of the LMC observed by \cite{Krishnarao2022} is not the primordial LMC CGM (i.e. CGM the LMC possessed at infall). Rather, this gas is likely a combination of new outflows from ongoing star formation in the LMC's disk as well as shocked CGM gas from the LMC's bow shock.

% As such, any potential ionized corona surrounding the LMC today must have been recently created, likely from a combination of accreted MW's CGM gas mass through LMC bow shock and ongoing stellar outflows from the LMC disk.

\begin{deluxetable*}{ccccccccccccccc}

\tabletypesize{\scriptsize}

\tablecaption{Visual classification of satellite positions relative to the LMC shock at a given time. All satellites that we find to have been near the shock at any point in the simulation are listed here. A label of $<10$ kpc indicates that a MW satellite is within 10 kpc of any point of the 3D shock boundary at that step in the simulation. \label{table:satellite_positions}}

\tablehead{
\colhead{Time} & \colhead{Aqu2} & \colhead{Car3} & \colhead{Hor1} & \colhead{Hor2} & \colhead{Hyi1} & \colhead{Hyd2} & \colhead{Phx2} & \colhead{Pis2} & \colhead{Ret2} & \colhead{Sag2} & \colhead{SMC} & \colhead{Tuc3} & \colhead{Tuc4} & \colhead{Tuc5} \\[-0.2cm]
\colhead{[Myr Ago]} & \colhead{} & \colhead{} & \colhead{} & \colhead{} & \colhead{} & \colhead{} & \colhead{} & \colhead{} & \colhead{} & \colhead{} & \colhead{} & \colhead{} & \colhead{}
}

\startdata
-500 &  & $<$10 kpc &  &  &  &  &  & $<$10 kpc &  &  &  &  &  &  \\
-400 &  & $<$10 kpc &  &  &  & $<$10 kpc & $<$10 kpc &  &  & In shock & &  &  &  \\
-300 &  & $<$10 kpc &  &  & In shock &  & In shock &  &  & In shock & &  &  &  \\
-200 & $<$10 kpc & In shock &  &  & In shock &  & $<$10 kpc &  &  & & $<$10 kpc  &  &  &  \\
-100 &  & $<$10 kpc &  &  & In shock &  &  &  &  & $<$10 kpc & In shock & & In shock & $<$10 kpc \\
Present &  &  & $<$10 kpc & $<$10 kpc & $<$10 kpc &  &  &  & In shock &  & In shock & $<$10 kpc &  &  \\ 
\enddata

\end{deluxetable*}

\subsection{Interaction of the LMC's Bow Shock with Satellites}

As the LMC's bow shock intersects with a significant volume of the ambient CGM, it is also possible that the bow shock has interacted with satellite galaxies associated with the LMC. To investigate which satellite galaxies may have interacted with the bow shock, we use orbital histories calculated following the methodology of \cite{Patel2020}, which includes the combined gravitational influence of the MW, LMC, and SMC. 

We adopt their low MW mass (virial mass of $1\times 10^{12} \, M_{\odot}$) gravitational potential and identical parameters. However, the MW center of mass is held fixed and does not move in response to the LMC's passage, since the MW is also fixed relative to the LMC in the orbit used in the \cite{Salem2015} simulations. Secondly, we model the LMC's gravitational potential as a Plummer sphere with a mass of $1.8 \times 10^{11} \, M_{\odot}$ and a Plummer scale length of $a=20$ kpc. The dynamical friction imparted by the MW and LMC remain unchanged. All satellites were treated identically to \cite{Patel2020}. They identified six satellite galaxies that may be dynamically associated to the LMC. These include six ultra-faint dwarfs (Phx2, Ret2, Car2, Car3, Hor1, Hyi1) and the SMC. We construct orbital histories using the line-of-sight velocities and distances as listed in \cite{Patel2020}. Proper motions are adopted from $Gaia$ eDR3 as reported in \citet{mcconnachie20}. 

We use the defined LMC+shock system at each of the six timesteps in Figure \ref{fig:shock_sat} along with the orbital histories of all ultra-faint and classical dwarf galaxies explored in \cite{Patel2020} to assess whether a given satellite is near or inside the shock at a given timestep. We flag satellites as ``In shock" if they are located between the 3D shock boundary and the LMC, and additionally identify galaxies that are within 10 kpc of the 3D shock boundary. The results of this visual classification are presented in Table \ref{table:satellite_positions}. Note that we have not explored the orbital uncertainties resulting from measurement errors on the 6D phase space properties of each galaxy. 

The LMC's companion galaxy, the SMC, has spent a considerable amount of its recent history inside the shock and near the boundary (see the star in Figure \ref{fig:shock_sat}). The SMC is also moving supersonically through the CGM: its speed of $\sim 250$ km/s \citep{Zivick2018} translates to a Mach number of $\sim$1.5, assuming the same model for the CGM as adopted in our simulations. Since the SMC is a gas rich galaxy, it should also produce a bow shock, but that shock will be weaker and smaller than that of the LMC. Interestingly, since the SMC is in orbit about the LMC, it is likely that the shocks of the LMC and the SMC themselves have been interacting over the past $\sim500$ Myr. The interaction between bow shocks could provide an additional source of heating to the CGM. This may be more generalizable in galaxy cluster environments.  

We find that Car3 and Hyi1 are the most likely other candidates for extended periods of interaction with the LMC shock; they appear to have spent $\sim500$ Myr of their orbits near the shock and have likely crossed the boundary at least once. Phx2, and Sag2 are also candidates for shock interaction given the amount of time ($\sim300$ Myr) they have spent near the LMC. Additionally, we find that Ret2 is inside the shock in the present day simulation slice (see Figure \ref{fig:shock_sat}).

While the jump in density due to the shock is modest, the physical extent of the shock is significantly larger than that of the LMC's gas disk. Interactions with this higher density gas may have had the effect of enhancing ram pressure on the gas in associated satellite galaxies, expediting the removal of gas.  Ultra-faint dwarfs are largely quenched during the epoch of reionization \citep{Brown2014}, although there may be some differences in quenching timescales for the LMC satellites \citep{Sacchi2021}. While these results imply that ultra-faint dwarfs would not retain significant amounts of star forming gas at late times, cosmological simulations illustrate that isolated ultra-faint dwarfs can retain significant reservoirs of ionized gas, and even neutral gas (depending on their halo mass), after reionization \citep{Jeon2017, Jeon2019}.  \cite{Emerick2016} use hydrodynamic simulations to illustrate that ram pressure stripping is not very efficient at removing gas from the inner regions of even ultra-faint dwarfs. Meaning that it is plausible that ultra-faint dwarfs may retain some gas reservoirs even while in orbit about the MW.  If ultra-faint dwarfs also cross the LMC bow shock boundary, they would experience higher ram pressure efficiency than average. This may help to remove all gas from the system, or maybe augment the ionization state of the gas inside the galaxy. 
 
The idea that ultra-faint dwarfs may interact with or be in proximity to the LMC's bow shock raises an interesting connection with indirect dark matter detection efforts. In particular, we find that the Ret2 ultra-faint dwarf is currently located inside the LMC's bow shock. Intriguingly, this galaxy has been found to harbor a gamma ray excess that has been potentially attributed as a signal of dark matter annihilation \citep{Hooper2015, Bonnivard2015, Geringer-Sameth2015, Drlica-Wagner2015}. While we do not find that the gamma ray emission from the shock itself will be substantial, it could help to augment any gamma ray signal that already exists in Ret2 owing to dark matter annihilation. Moreover, if Ret2 is interacting with the shock and harbors any residual gas, the gas could be sufficiently excited to emit gamma rays. This scenario may help to explain why the signal is particularly strong in Ret2, versus other similarly massive ultra-faint dwarfs.

We suggest that the LMC bow shock may be an intriguing complication to the interpretation of any detection of gamma ray excess in dwarf galaxies that are in its proximity, making our study of the geometry of the shock a critical tool to understand all astrophysical origins for a gamma ray excess. 

% \newpage

\section{Conclusions}

Using a simulation of the LMC's gas disk over the past 500 Myr \citep{Salem2015}, we characterize the size and structure of the bow shock that is predicted to accompany the LMC as it falls into the MW potential at supersonic speeds ($M\sim2.1$). We find that:

\begin{itemize}
    \item The MW CGM gas leading the LMC is expected to jump in density and temperature by a factor of 2-3, matching theoretical predictions (see Figure \ref{fig:shock_conditions}). We characterize the shape of the shock and find that it is asymmetric due to the mismatch in orientation between the LMC's disk and velocity vector (see Figure \ref{fig:vel_slice}).
    
    \item We predict that the distance to the shock front along the direction of the LMC velocity vector should be $\sim6.7$ kpc with a sharply defined discontinuity. The shock's approximate transverse diameter should be $\sim30$ kpc at present day.

    \item The shock swept over a large fraction of the CGM ($\sim8\%$ of the southern sky at r$<$140 kpc, see Figure \ref{fig:shock_sat}). The infall of massive satellites may present an important and understudied dynamical mixing process in the CGM. 

    \item The morphology of the predicted shock aligns very well with observed H$\alpha$ emission that extends along the LMC leading edge and moves at speeds consistent with the systemic speed of the LMC (see Figure 7 in \citealt{Smart2023}). We propose that this occurs owing to interactions between the shocked MW CGM gas and the cold clouds surrounding the LMC, making this emission a signature of the existence of the bow shock.

    \item Given the ram pressure stripping required to truncate the LMC gas disk \cite{Salem2015}, our results suggests that the observed increase in ionized gas near the LMC \citep[e.g.][]{Krishnarao2022}, is newly formed from stellar outflows or shocked MW CGM gas, rather than being a primordial, pre-infall LMC CGM.
    
    \item The SMC has spent the past $\sim200$ Myr inside or near the shock. Because it is moving supersonically, it is very likely that there have been shock-shock interactions between the LMC and SMC shocks that could act as source of CGM heating. 
    
    \item Many other satellite galaxies may have interacted with the shock over the past 500 Myr. In particular, we find that Car3, Hyi1, Phx2, and Sag2 are likely candidates for having interacted with the shock, which might have helped to remove gas from these systems. We further find that at present day, the Ret2 ultra-faint dwarf is in proximity to the shock, which may complicate efforts to understand the origin of any detected gamma ray excess.

\end{itemize}

The presence of a massive satellite with a bow shock in the CGM of our own Galaxy underscores the importance of studying the dynamics of massive satellites in addition to inflows/outflows to understand the properties of the multi-phase CGM. In particular, we posit that the accretion of massive satellites is an important dynamical process that can aid in mixing in the CGM. Given that most MW-mass galaxies have accreted an LMC-mass subhalo at some point in the past \citep{Stewart2008}, this work may have relevance for understanding the CGM of other MW analogs \citep[e.g.,][]{Tumlinson2011, Lehner2020}. More generally, this study has relevance for a wide range of environments where galaxies are moving at high speeds through a gaseous medium, such as cluster jellyfish galaxies \citep[e.g.,][]{Poggianti2017, Werle2022}. 

This study further illustrates how the LMC can change the kinematic properties of the MW's CGM, forcing CGM gas to move at speeds similar to its own systemic velocity. This result has important consequences for the interpretation of the origin of ionized gas associated with the LMC/SMC system. This gas has thus far been attributed to gas removed from the Magellanic Clouds, however, we argued that this gas could simply be CGM gas from the MW with kinematics that have been altered by the passage of the LMC. 

\acknowledgements

Support for this work was provided by The Brinson Foundation through a Brinson Prize Fellowship grant. GB acknowledges support from NSF CAREER award AST-1941096. EP acknowledges financial support provided by NASA through grant number HST-GO-16628. Support for this work was also provided by NASA through the NASA Hubble Fellowship grant \# HST-HF2-51540.001-A awarded by the Space Telescope Science Institute, which is operated by the Association of Universities for Research in Astronomy, Incorporated, under NASA contract NAS5-26555.

We thank Yumi Choi for sharing her orthographic projection code, which makes it possible to directly compare observed H$\alpha$ emission to the simulation in Figure 5. 
% We thank Munier Salem for his help in constructing the LMC LOS projections in Figure 1 and Figure 5, and for running the simulations that enabled this work. 
We also thank Xiawei Wang for assisting with the calculations related to the observability of the shock. Additionally, we thank Jess Werk and Mary Putman for providing early comments on the paper. Finally, we thank Brianna Smart for helping us acquire the WHAM data for the generation of Figure 5.

\software{Astropy \citep{astropy:2022},
Matplotlib \citep{Hunter:2007}, yt \citep{yt}, Enzo \cite{Bryan2014}}

\bibliography{LMC_Bow_Shock}

\end{document}